\documentclass[aps,prl,twocolumn,amsmath,amssymb,nofootinbib,
  superscriptaddress]{revtex4-1}

\usepackage{graphicx}
\usepackage{bm}
\usepackage[colorlinks]{hyperref}
\usepackage[usenames,svgnames]{xcolor}
\usepackage{makecell}
\usepackage[normalem]{ulem}

\usepackage{algorithm}
\usepackage{algorithmicx}
\usepackage{algpseudocode}
\usepackage{makecell}

\newcommand{\hhl}[1]{{\color{black}{#1}}}

\newcommand{\sfacost}{\mathcal{C}_{SFA}}

\providecommand{\U}[1]{\protect\rule{.1in}{.1in}}

\begin{document}
\title{How to Design a Classically Difficult Random Quantum Circuit for Quantum Computational Advantage Experiments}

\author{He-Liang Huang}
\thanks{These authors contribute equally}
\email{quanhhl@ustc.edu.cn}
\affiliation{Henan Key Laboratory of Quantum Information and Cryptography, Zhengzhou, Henan 450000, China}
\affiliation{Hefei National Research Center for Physical Sciences at the Microscale and School of Physical Sciences, University of Science and Technology of China, Hefei 230026, China}
\affiliation{Shanghai Research Center for Quantum Science and CAS Center for Excellence in Quantum Information and Quantum Physics, University of Science and Technology of China, Shanghai 201315, China}

\author{Youwei Zhao}
\thanks{These authors contribute equally}
\affiliation{Hefei National Research Center for Physical Sciences at the Microscale and School of Physical Sciences, University of Science and Technology of China, Hefei 230026, China}
\affiliation{Shanghai Research Center for Quantum Science and CAS Center for Excellence in Quantum Information and Quantum Physics, University of Science and Technology of China, Shanghai 201315, China}

\author{Chu Guo}
\email{guochu604b@gmail.com}
\affiliation{Key Laboratory of Low-Dimensional Quantum Structures and Quantum Control of Ministry of Education, Department of Physics and Synergetic Innovation Center for Quantum Effects and Applications, Hunan Normal University, Changsha 410081, China}

\begin{abstract}

Quantum computational advantage is a critical milestone for near-term quantum technologies and an essential step towards building practical quantum computers. Recent successful demonstrations of quantum computational advantage owe much to specifically designed random quantum circuit (RQC) protocols that enable hardware-friendly implementation and, more importantly, pose great challenges for classical simulation. Here, we report the automated protocol design approach used for finding the optimal RQC in the \textit{Zuchongzhi} quantum computational advantage experiment [Phys. Rev. Lett. 127 (18), 180501 (2021)]. Without a carefully designed protocol, the classical simulation cost of the \textit{Zuchongzhi}'s 56-qubit 20-cycle RQC experiment would not be considerably higher than Google's 53-qubit 20-cycle experiment, even though more qubits are involved. For Google's latest RQC experiment using $70$ qubits and $24$ cycles [arXiv:2304.11119 (2023)], we estimate that its classical simulation cost can be increased by at least one order of magnitude using our approach. The proposed method can be applied to generic planar quantum processor architectures and addresses realistic imperfections such as processor defects, underpinning quantum computational advantage experiments in future generations of quantum processors.

\end{abstract}
\date{\today}

\maketitle

A primary pursuit in the noisy intermediate-scale quantum (NISQ) computing era is to fully understand and unlock the computational power of near-term quantum devices~\cite{huang2023near,huang2020superconducting,RevModPhys.94.015004,kjaergaard2020superconducting,haffner2008quantum,krantz2019quantum,bruzewicz2019trapped,wang2020integrated,slussarenko2019photonic,flamini2018photonic,cerezo2021variational}. In this stage, a crucial question to ask is whether noisy quantum computers, which are incapable of error correction, can outperform state-of-the-art supercomputers for solving certain tasks, i.e., realize quantum computational advantage (QCA)~\cite{preskill2012quantum}. 
This intriguing question has sparked a continuing effort in developing more powerful quantum computers, such as Sycamore~\cite{AruteMartinisQuantumSupremacy2019,Google2023b}, \textit{Zuchongzhi}~\cite{WuPan2021,ZhuPan2021}, \textit{Jiuzhang}~\cite{zhong2020quantum,PhysRevLett.127.180502,deng2023gaussian,PhysRevLett.130.190601} and Borealis~\cite{madsen2022quantum}, as well as more efficient classical simulators~\cite{GuoWu2019,VillalongaMandra2018,VillalongaMandra2019,GrayKourtis2020,HuangChen2021,GuoHuang2021,PanZhang2021,LiuChen2021,PanZhang2021b,ChenYang2022,LiuLiu2022,xu2023herculean}. However, with continuous improvements from both sides, the demonstration of QCA should not be considered as a single-shot achievement, but as a long-term competitive process of quantum and classical devices (and algorithms) in the NISQ era.

Specially designed protocols play an important part in the competitive game of demonstrating QCA; after all, current quantum hardwares are not equipped to achieve advantages on a wide range of problems. One leading protocol is random circuit sampling (RCS)~\cite{boixo2018characterizing}: the task of sampling from the output distribution of random quantum circuits (RQCs). RCS has two excellent properties for QCA experiments: 1)~The RQCs, which consist of only nearest-neighbor two-qubit gates on a 2D grid and single-qubit gates randomly selected from a small gate set, are suitable choices for superconducting quantum processors; 2)~Meanwhile, RCS satisfies an average-case \#P-hardness~\cite{bouland2019complexity,mi2021information}, which is commonly viewed as hard to classically simulate. Beyond that, RCS has been shown to be a powerful benchmarking primitive for extracting the total amount of quantum noise in a multi-qubit quantum system~\cite{liu2021benchmarking}.

An easily overlooked issue, which however has to be seriously considered in attempts to use RCS to establish QCA, is that the structure of a RQC is something that needs to be carefully designed to enlarge the gap between quantum computing and classical simulation. Specifically, the patterns of the two-qubit gate layers have a crucial effect on the classical simulation cost. For example, if \textit{Zuchongzhi}'s 56-qubit 20-cycle RCS experiment~\cite{WuPan2021} strictly follows the same two-qubit gate patterns as Sycamore's 53-qubit 20-cycle RCS experiment~\cite{AruteMartinisQuantumSupremacy2019}, then it can only achieve very minor QCA beyond that of Sycamore's 53-qubit 20-cycle RCS~\cite{AruteMartinisQuantumSupremacy2019}, even though more qubits are used (see the analysis in the `\textit{Numerical results}' section). Thus, the RQC needs to be customized for each quantum processor, for the purpose of QCA experiments. 

In this work, we present the method we have used for designing the optimal RQCs in the \textit{Zuchongzhi} QCA experiment. 
Our method mainly contains two ingredients: a) efficiently estimating the classical simulation cost for a specific RQC with fixed two-qubit gate pattern; b) efficiently exhausting all the possible two-qubit gate patterns to find the ones with the highest classical simulation cost. Our method is general for Sycamore-like quantum processors with adjustable number of qubits and circuit depths, and could be straightforwardly generalized to other planar architectures. It can also take into account of hardware imperfections such as defective qubits. 
As application, we estimate that when applied to Google's latest QCA experiment using $70$ qubits and $24$ cycles~\cite{Google2023b}, our method could further increase the classical simulation cost by at least one order of magnitude.
These features demonstrate that our method could be an universal tool for designing future QCA experiments.

\begin{figure}
  \includegraphics[width=0.95\columnwidth]{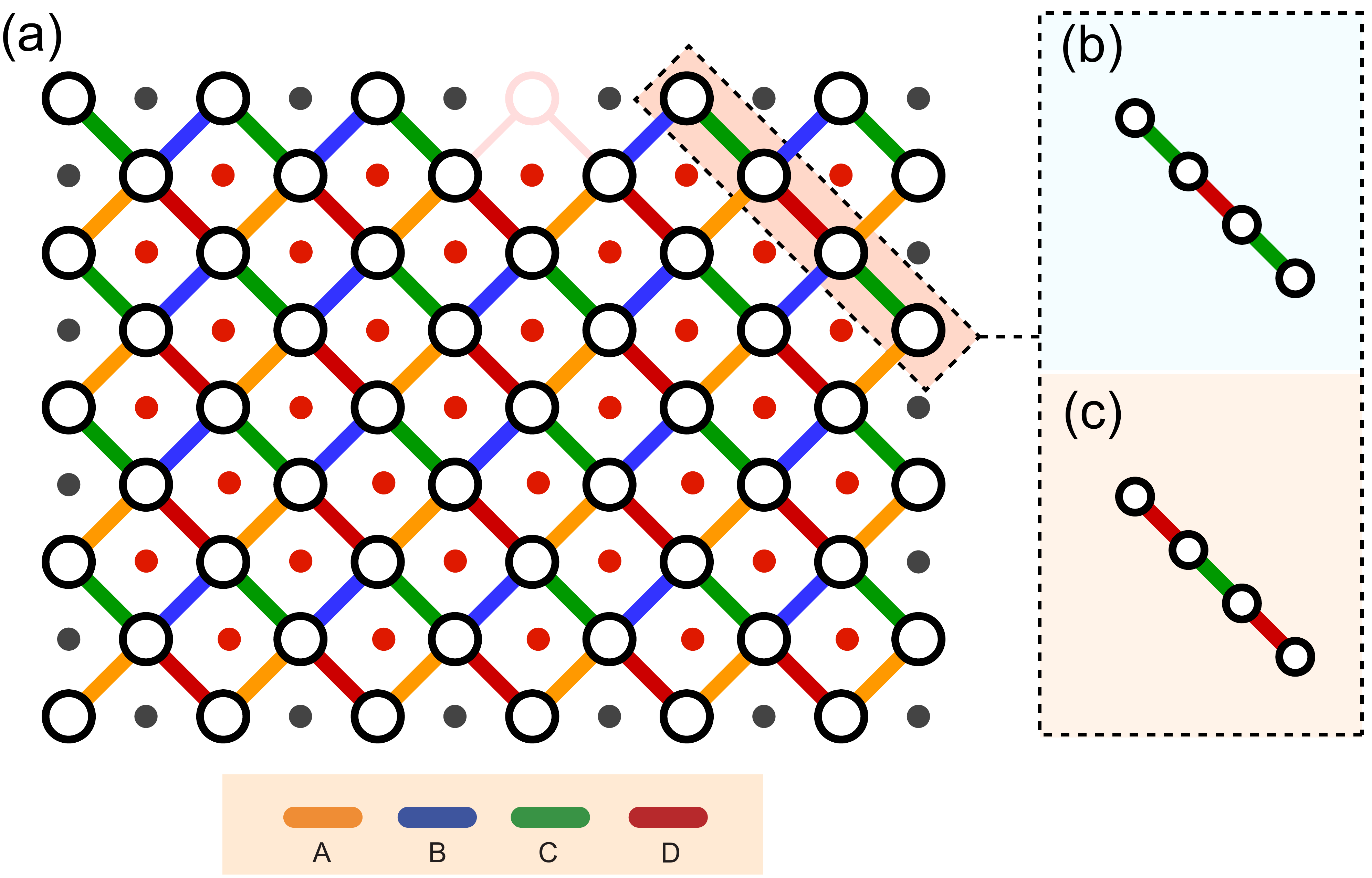} 
  \caption{\textbf{The Sycamore quantum processor and the RQC.} (a) The empty circles represent the qubits, and the shaded empty circle represents the defective qubit. The edge between a nearest-neighbour pair of qubits represents a two-qubit gate operation, and the edges with the same color represent the two-qubit gates which are performed simultaneously. (b) and (c) are two possible configurations for the two-qubit gate patterns \textit{C} and \textit{D} in the boxed region.
    }
    \label{fig:fig1}
\end{figure}

\textit{RQC on Sycamore-like quantum processors.} The architecture of the Sycamore quantum processor as well as the two-qubit gate patterns used for RCS experiments are compactly shown in Fig.~\ref{fig:fig1}(a). 
The processor architecture could either be extended in the horizontal or the vertical directions so as to accommodate more qubits. The edge on a nearest-neighbour pair of qubits (a bond) represents a two-qubit gate
, and the edges with the same color represent the two-qubit gates which are performed in parallel, referred to as one \textit{cycle}. A full layer of random single-qubit gates are assumed to be performed between two successive cycles (and at the beginning and end of the circuit). We can see that the four cycles in Fig.~\ref{fig:fig1}(a), denoted as $ABCD$, ensure that a two-qubit gate has been applied on each bond (thus all the qubits can be entangled).
To build RQCs with more than four cycles, we will simply repeat the pattern $ABCDCDAB$ as in Ref.~\cite{AruteMartinisQuantumSupremacy2019}, for example, an RQC with $12$ cycles will be $ABCDCDABABCD$.

It should be noted that in a RQC for a given processor architecture, the two-qubit gate pattern is not unique. For example, for the two-qubit gate patterns \textit{C} and \textit{D} in the shaded region of Fig.~\ref{fig:fig1}(a), they can either be aligned as in Fig.~\ref{fig:fig1}(b) or Fig.~\ref{fig:fig1}(c) without affecting the parallel execution. The overall quantum fidelities of different designs of an $N$-qubit $d$-cycle RQC are roughly the same (since the total number of gates is almost unchanged)
, but their classical simulation cost varies significantly. For example, Ref.~\cite{AruteMartinisQuantumSupremacy2019} proposed 
a largely simplified RQC by simply rearranging the two-qubit gates in $ABCD$ to enable efficient classical verification.  
It is not obvious to obtain an optimal choice of the RQC to maximize the classical simulation cost. This design question is solved in this work with the two techniques shown in the following.

\textit{Efficiently exhausting all the possible RQCs.} 
The first step to find an optimal RQC is to efficiently exhaust all the possible two-qubit gate patterns, $ABCD$. 
These four patterns are not independent. We can see that $B$ is fixed once $A$ is fixed, and the same holds for $D$ and $C$. The different patterns of $A$ and $C$ can be counted as follows. To count all the different possibilities of $A$, we can view the lattice in the $\nearrow$ direction. In Fig.~\ref{fig:fig1}(a) there are $9$ rows in the $\nearrow$ direction, and in each row, one can choose to perform two-qubit gate operations on either the even (represented as $0$) or the odd (represented as $1$) bonds. With this convention the pattern $A$ shown in Fig.~\ref{fig:fig1}(a) can be uniquely specified by the bitstring $111111111$, viewed from top left to bottom right. Similarly, the pattern $C$ can be represented as $000000000$, viewed from bottom left to top right. 
Therefore if we denote $m$ as the number of rows in the $\nearrow$ direction, and $n$ as the number of rows in the $\searrow$ direction, then the total number of choices for the first $ABCD$ cycles is $2\times 2^m\times 2^n = 2^{m+n+1}$, where the factor of $2$ is due to the two different choices by interchanging $AB$ and $CD$. With this notation, we can see that the RQCs used in Google's QCA experiments have chosen $A$ with all $1$s and $C$ with all $0$s~\cite{AruteMartinisQuantumSupremacy2019,Google2023b}.

\textit{Efficiently estimating the classical simulation cost of a given RQC.}
Once we have identified all the possible RQCs, the remaining task is to estimate the classical simulation cost for each of them, based on which we can select the optimal one with the highest cost. The classical simulation cost of an RQC is highly dependent on the classical simulation algorithm used. 
Currently the tensor network contraction (TNC) method has demonstrated itself to be the most efficient classical simulation algorithm~\cite{GuoWu2019,VillalongaMandra2018,VillalongaMandra2019,GrayKourtis2020,HuangChen2021,GuoHuang2021,PanZhang2021,LiuChen2021,PanZhang2021b,ChenYang2022,LiuLiu2022} for large-scale RQCs. The classical simulation cost of the TNC algorithm relies crucially on a heuristic tensor network contraction order (TNCO), while searching for a good TNCO for a moderate RQC with $N \approx 60$, $d=20$ could often take hours of time. Taking into account that there could be more than $10^5$ different RQCs, it would be impractical to use the TNC algorithm to estimate the classical simulation cost. 

We choose to use the Schr$\ddot{\text{o}}$dinger-Feymann algorithm (SFA) instead to estimate the classical simulation cost~\cite{AruteMartinisQuantumSupremacy2019,markov2018quantum}. The main idea of SFA is to divide the whole system into two subsystems, and store two quantum states corresponding to these two subsystems brute-forcely as state vectors. The cost of SFA is determined by the entanglement generated between the two subsystems. Specifically, the gate operations which only act on one subsystem can be simply applied on the subsystem. While the two-qubit gate $G_{i, j}$ that acts on both subsystems (cross gate) has to be first split into the summation $G_{i, j} = \sum_1^{\chi} L_i^s \otimes R_j^s$ ($2\leq\chi\leq 4$ is the Schmidt rank of the two-qubit gate, which is $4$ for Fsim gate~\cite{AruteMartinisQuantumSupremacy2019}), and then apply each pair $L_i^s$ and $R_j^s$ onto the corresponding subsystems. SFA overcomes the memory issue of simulating large quantum systems since only the two quantum states for the two subsystems need to be stored, while the disadvantage is that $\chi$ branches are generated whenever a cross gate is encountered and the computational cost scales as $\chi^{n_c}$ for $n_c$ cross gates (see Ref.~\cite{markov2018quantum} for more details of SFA). 

\begin{algorithm}[H]
\caption{Depth first search (DFS) algorithm to find all possible bipartition. We denote the set of all boundary sites on the dual lattice as $\mathcal{B}$, and the threshold for the number of edges in a path as $E^{\ast}$. A path of $m$ edges is specified by $m+1$ successive sites on the dual lattice. } \label{alg:paths}
\begin{algorithmic}[1]
\State Initialize: $\mathcal{P} = \{\}$
\For{$q_0 \in \mathcal{B} $ }
\State ${\rm DFS}(\{q_0\})$
\EndFor
\Procedure{DFS}{$p$}
\If{${\rm Length}(p) > E^{\ast}$}
  \State Return
\EndIf
\For{$q \in {\rm Neighbours}(p[{\rm end}])$ and $q \notin p$}
\State $p^{\prime} = \left[p; q \right]$ 
\If{$q\in \mathcal{B}$}
\State $\mathcal{P} = \left[\mathcal{P}; p^{\prime} \right]$
\Else
\State ${\rm DFS}(p^{\prime})$
\EndIf
\EndFor
\EndProcedure
\end{algorithmic}
\end{algorithm}

To estimate the computational cost of SFA, one needs to find an optimal way to divide the system into two subsystems, such that the total number of cross gates $n_c$ is minimized. 
The brute force way is to identify all the possible bipartition of the lattice, and count $n_c$ for each of them. 
At first sight the number of possible bipartition seems to grow exponentially, however, by imposing two reasonable constraints this number can be greatly reduced. 
To see this we define a dual lattice whose lattice sites are the centers of the original lattice, as indicated by the solid circles in Fig.~\ref{fig:fig1}(a). The boundary sites of the dual lattice sites are marked black while the interior sites are marked red. 
Now we can see that a bipartition of the original lattice corresponds to draw a connected path on the dual lattice, which can either be an open path whose endpoints are located at the boundary sites of the dual lattice, or a closed path. We can also see that $n_c$ is closely related to the number of edges $E$ in the path as $n_c = Ed/4$. Therefore to minimize $n_c$ one should generally find an optimal path with minimal $E$.
The first constraint is that only open paths should be chosen, since it can be seen that $E$ for a closed path will be much larger than that for a good open path. Second, for Sycamore-like lattice one can draw a path from top down or from left to right, and the number of edges for this path, denoted as $E^{\ast}$, should not be smaller than $E$ for the optimal path. With this observation, the open paths with $E > E^{\ast}$ are thrown away. The pseudocode to collect all the open paths are shown in Algorithm.~\ref{alg:paths}, where the function ${\rm Neighbours}(q)$ returns all the neighbouring sites of site $q$, and the function $\left[p; q\right]$ on a set $p$ and an element $q$ returns a new set with all elements of $p$, and adds $q$ as the last element.

\begin{table*}[!htbp]
  \begin{center}
    \caption{\hhl{\textbf{Runtime for evaluating the SFA and TNC costs for RQCs of different sizes}, denoted as \textit{SFA runtime} and \textit{TNC runtime}, respectively. The evaluation of SFA cost utilized only a single core, while the evaluation of TNC cost utilized $56$ cores. For RQCs of sizes 56-qubit 20-cycle, 60-qubit 24-cycle, and 70-qubit 24-cycle, there are $2^{19}$, $2^{20}$, and $2^{21}$ candidate circuits to be evaluated, respectively. We provide the time required for evaluating a single circuit and evaluating all circuits separately. Runtime ``$s$" and ``$d$" represent seconds and days.
    }}
    \label{tab:tab1}
    \begin{tabular}{c|c|cc|cc}
    \hline
    \hline
    RQC sizes & Number of circuits & \makecell{SFA runtime ($s$) \\for a circuit \\ (one core)} & \makecell{SFA runtime ($d$) \\for all circuits\\ (one core)} & \makecell{TNC runtime ($s$) \\for a circuit\\ (56 cores)} &\makecell{TNC runtime ($d$) \\for all circuit\\ (56 cores)}\\
    \hline
    $56\times 20$ & $2^{19}$  & $0.040$  & $0.24$ & $3512$ & $21311$ \\
    $60\times 24$ & $2^{20}$  & $0.108$  & $1.31$ & $6414$ & $77842$  \\
    $70\times 24$ & $2^{21}$  & $0.063$  & $1.52$ & $10668$ & $258940$ \\
  \hline
    \end{tabular}
  \end{center}
\end{table*}

To evaluate the computational cost of SFA for a given bipartition, denoted as $\sfacost$, there are several additional complexities to be taken into account. First, the number of wedges, $n_{wedge}$ where two successive two-qubit gates act on a common qubit, and the number of $DCD$ formations, $n_{DCD}$, are going to reduce the effective number of cross gates (See Ref.~\cite{AruteMartinisQuantumSupremacy2019} and SM for details of these two formations). Second,  in the first and final cycles, the Fsim gate can be simplified to \textit{cphase} with $\chi=2$ only, thus the contributions of the cross gates in the first ($n_{st}$) and final ($n_{end}$) cycles will be reduced by half. Overall, $\sfacost$ for a given bipartition is estimated as
\begin{align}\label{eq:sfacost}
\sfacost = 4^{n_c - n_{wedge} - n_{DCD} - \frac{n_{st} + n_{end}}{2} }\times (2^{n_1}+2^{n_2}).
\end{align}
where $n_1$ and $n_2$ are the number of qubits in the two subsystems. Given a fixed total number of qubits $N$, the contribution of the last term in Eq.(\ref{eq:sfacost}) will be much larger in the `imbalanced' situation with a large $|n_1-n_2|$ compared to the balanced case with $n_1\approx n_2$. Moreover if one subsystem is too large we will again have memory issue. Therefore we also set a threshold $n^{\ast}$ in Algorithm.~\ref{alg:paths} such that any path with $|n_1-n_2| > n^{\ast}$ will be neglected. With Algorithm.~\ref{alg:paths} and Eq.(\ref{eq:sfacost}), we could find the minimal $\sfacost$ for a given RQC.

Table.~\ref{tab:tab1} shows the time required to estimate the computational cost of SFA and TNC for RQCs of different sizes. Here we have used package cotengra~\cite{GrayKourtis2020} to search for the optimal TNCO. Throughout this work, the major hyperparameters we set for contengra are as follows. The maximally allowed intermediate tensor size is set to be $2^{30}$. The parameter `max\_repeats' is set to be $1000$ and the method `kahypar' is used for optimization. 
It is clear from Table.~\ref{tab:tab1} that the time taken to evaluate the SFA complexity is much lower and the advantage becomes more apparent as the size of the RQC increases. 
\hhl{Importantly, despite that we only use one core to evaluate the SFA cost but use $56$ cores to evaluate the TNC cost, to exhaust all the possible 70-qubit 20-cycle RQCs, the former takes 1.52 days, while the latter takes an unacceptable 258,940 days. Even with the world's most powerful supercomputer, Frontier, which has 8.73 million computer cores, the total time to estimate the TNC costs for all 70-qubit 24-cycle RQCs would be $258,940~\text{days} \times 56 / (8.73 \times 10^6) = 1.6~\text{days}$, which is indeed staggering. In addition, the search for the optimal TNCO is heuristic, meaning that a single evaluation of the TNC cost is often not precise enough (see SM for details), but performing multiple evaluations would be even more time-consuming.} 
Therefore, using TNC as a benchmark for designing RQCs is not practical, whereas our method can be well extended for RQCs with even more qubits and cycles.

\begin{figure}[tbp!]
  \includegraphics[width=1\columnwidth]{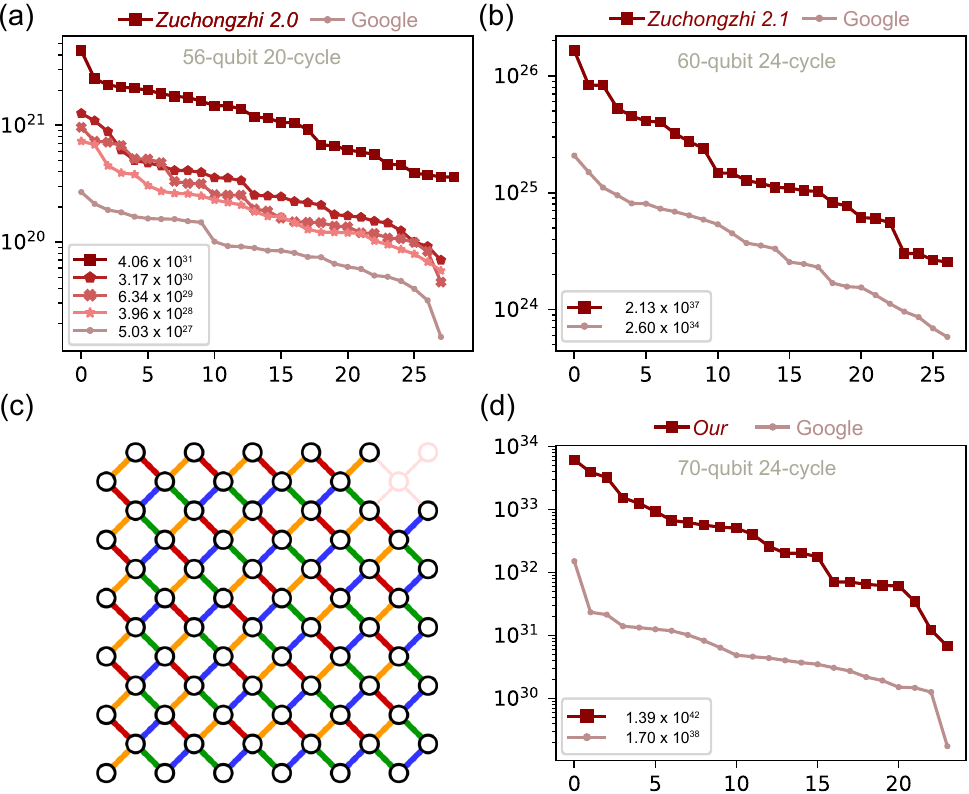} 
  \caption{\textbf{Correspondence between the TNC costs and the SFA costs.} The TNC costs for RQCs with (a) $56$ qubits and $20$ cycles, (b) $60$ qubits and $24$ cycles and (d) $70$ qubits and $24$ cycles, where for each RQC we have run cotengra~\cite{GrayKourtis2020} for more than $20$ instances with the costs sorted in descending order. The corresponding SFA cost is shown in the legend. `Google' represents RQCs generated directly from Google's experiments~\cite{AruteMartinisQuantumSupremacy2019,Google2023b}, and `\textit{Zuchongzhi} 2.0'~\cite{WuPan2021}, `\textit{Zuchongzhi} 2.1'~\cite{ZhuPan2021}, and 'our' represent RQCs designed using our method. 
  (c) The optimal RQC designed by our method for Google's latest 70-qubit 24-cycle QCA experiment~\cite{Google2023b}.
    }
    \label{fig:fig2}
\end{figure}

\textit{Numerical results.} A crucial question is whether the RQC obtained by our approach is also challenging for other classical simulation algorithms, in particular the TNC algorithm. To demonstrate the effectiveness of our approach, we generate five RQCs with 56 qubits and 20 cycles for \textit{Zuchongzhi 2.0} processor~\cite{WuPan2021}, each with different SFA complexities (see SM for the two-qubit gate patterns of these circuits). The circuit with the highest SFA complexity, which is also the one used in the \textit{Zuchongzhi 2.0} experiment, is denoted as 56-20-\textit{Zuchongzhi 2.0}. The circuit with the lowest complexity is directly adopted from Google's experiment~\cite{AruteMartinisQuantumSupremacy2019,Google2023b} and referred to as 56-20-Google (similar notations are used in the following). The complexities of the other three circuits fall between these two. 
As shown in Fig.~\ref{fig:fig2}(a), it is evident that circuits with higher SFA cost also have higher TNC costs \hhl{(In the SM, we show that RQCs with higher SFA costs often exhibit faster entropy growth, which further underscores their difficulty for classical simulation)}. Table.~\ref{tab:tab2} gives the TNC costs of computing a noisy sample from 56-20-\textit{Zuchongzhi 2.0}~\cite{WuPan2021}, 56-20-Google, and 53-20-Google~\cite{AruteMartinisQuantumSupremacy2019}. Without using our method, the simulation cost for 56-20-Google is only higher than 53-20-Google by less than three times. However, our designed 56-20--\textit{Zuchongzhi 2.0} exhibits a classical simulation cost that is nearly two orders of magnitude higher than 53-20-Google. 
In Fig.~\ref{fig:fig2}(b) we further compare the TNC costs of the RQC designed with our method to those of Google's RQC, for the 60-qubit 24-cycle QCA experiment performed by \textit{Zuchongzhi 2.1}~\cite{ZhuPan2021}, which confirms the superior classical simulation cost of our design. 


\begin{table}[!tb]
  \begin{center}
    \caption{\textbf{Estimated classical simulation costs for different RQCs.} The second column shows the XEB fidelities of the \textit{Zuchongzhi 2.0}.~\cite{WuPan2021} and Sycamore~\cite{AruteMartinisQuantumSupremacy2019} QCA experiments. The third and forth columns show the FLOPS required for classically
    calculating a perfect and noisy sample from the RQC, respectively.}
    \begin{tabular}{l|ccc}
    \hline
    \hline
    RQC sizes & Fidelity & 1 perfect amp. & 1 noisy amp.  \\
    \hline
    $56\times 20 $ (\textit{Zuchongzhi})& $0.0662\%$  & $3.61 \times 10^{20}$ & $2.2 \times 10^{17}$\\
    $56\times 20$ (Google)& $0.0662\%$ & $1.54 \times 10^{19}$ & $1.02 \times 10^{16}$\\
    $53\times 20$ (Google)& $0.224\%$  & $1.63 \times 10^{18}$  & $3.65 \times 10^{15}$\\
  \hline
    \end{tabular}
    \label{tab:tab2}
  \end{center}
\end{table}

As application, we apply our method to search for the optimal RQC for the latest Google's QCA experiment with $70$ qubits and $24$ cycles~\cite{Google2023b}. The optimal RQC we found is shown in Fig.~\ref{fig:fig2}(c). 
And the TNC costs for our RQC and Google's RQC are shown in Fig.~\ref{fig:fig2}(d).
We can see that the lowest TNC costs of our RQC and Google's RQC are estimated as $6.7\times 10^{30}$ and $1.7 \times 10^{29}$ respectively (we note that these costs are both higher than the value reported in Ref.~\cite{Google2023b}, possibly due to different hyperparameters used or improved ways of searching for optimal TNCOs in Ref.~\cite{Google2023b} compared to cotengra). This means that if our RQC is used in this experiment, the computational time for classical simulation on Frontier supercomputer would increase from the original $4.68 \times 10$ years to $1.84 \times 10^3$ years (about 40 times).
\begin{acknowledgments}
\textit{Conclusion.} 
Both RCS and Boson sampling~\cite{aaronson2011computational} are elaborate protocols designed to maximize the computational power of quantum computing and overwhelm classical computing. Within a decade, the classical and quantum rivalry will likely continue unabated. 
By designing a classically difficult RQC using our method, at least one order of magnitude gain in classical simulation cost is achieved compared to Google's original RQC. This improvement is of great importance for the near-term development of quantum computing hardware. On the one hand, both quantum and classical hardware face significant challenges in performance enhancement, by improving the design of the RQC, we can enlarge the advantages of quantum computing without imposing new requirements on the quantum hardware. On the other hand, by changing the circuit structure, RQC with higher classical simulation costs can be obtained, the underlying reason perhaps being the faster growth of quantum entanglement, among others. In the future, by understanding this phenomenon and the physics behind it, it may help us to explore practical applications with quantum advantage experiments.

\hhl{In addition, it is worth noting that while our designed RQCs are challenging to simulate for current state-of-the-art TNC algorithms, it does not imply that they are difficult to simulate for all classical simulation methods, as classical simulation methods continue to improve. However, our entire method is a framework that can be compatible with other classical algorithms for accessing the simulation cost. In the future, if more powerful classical algorithms become available, our framework can be easily adapted to design RQCs based on these algorithms.}

H.-L. H. is supported by the Youth Talent Lifting Project (Grant No. 2020-JCJQ-QT-030), National Natural Science Foundation of China (Grants Nos. 11905294, 12274464). C. G. is supported by the Open Research Fund from State Key Laboratory of High Performance Computing of China (Grant No. 202201-00).

\textit{Note added.--All code will be open source after publication~\cite{RQC_Github}.}

\end{acknowledgments}

\bibliographystyle{apsrev4-1}
\bibliography{references}

\end{document}


\title{Supplemental Material for \\ ``How to Design a Classically Difficult Random Quantum Circuit for Quantum Computational Advantage Experiments''}

\author{He-Liang Huang}
\thanks{These authors contribute equally}
\email{quanhhl@ustc.edu.cn}
\affiliation{Henan Key Laboratory of Quantum Information and Cryptography, Zhengzhou, Henan 450000, China}
\affiliation{Hefei National Research Center for Physical Sciences at the Microscale and School of Physical Sciences, University of Science and Technology of China, Hefei 230026, China}
\affiliation{Shanghai Research Center for Quantum Science and CAS Center for Excellence in Quantum Information and Quantum Physics, University of Science and Technology of China, Shanghai 201315, China}

\author{Youwei Zhao}
\thanks{These authors contribute equally}
\affiliation{Hefei National Research Center for Physical Sciences at the Microscale and School of Physical Sciences, University of Science and Technology of China, Hefei 230026, China}
\affiliation{Shanghai Research Center for Quantum Science and CAS Center for Excellence in Quantum Information and Quantum Physics, University of Science and Technology of China, Shanghai 201315, China}

\author{Chu Guo}
\email{guochu604b@gmail.com}
\affiliation{Key Laboratory of Low-Dimensional Quantum Structures and Quantum Control of Ministry of Education, Department of Physics and Synergetic Innovation Center for Quantum Effects and Applications, Hunan Normal University, Changsha 410081, China}

\date{\today}

\maketitle

\setcounter{section}{0}
\renewcommand{\thefigure}{S\arabic{figure}}	
\renewcommand{\thetable}{S\arabic{table}}	
\renewcommand{\theequation}{S\arabic{equation}}	
\setcounter{figure}{0}
\setcounter{table}{0}
\setcounter{equation}{0}

\section{I. Identifying wedge and DCD formations during SFA complexity evaluation}

The wedge and DCD formations are going to reduce the bipartition entanglement, as pointed out in Ref.~\cite{AruteMartinisQuantumSupremacy2019}. Therefore they should be taken into account when estimating the classical simulation cost of SFA (they could also reduce they classical simulation cost of the projected entangled pair state based classical simulator as shown in Ref.~\cite{GuoHuang2021}).

\begin{figure}[!htbp]
\begin{center}
\includegraphics[width=\linewidth]{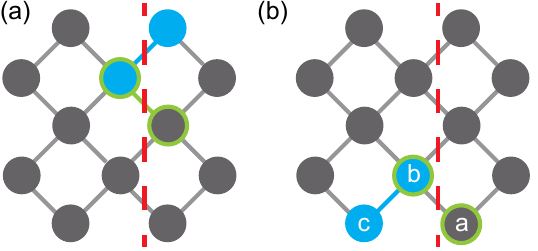}
\end{center}
\caption{\textbf{Wedge and DCD formations.} (a) The gate highlighted in blue and green on the cut line are crosspartition gates. A wedge is formed when two consecutive crosspartition gates share a qubit. (b) DCD formation.
\label{wedge}}
\end{figure}

1) \textit{Wedge formation}. As shown in Fig.~\ref{wedge} (a), a wedge formation is formed when two consecutive cross-partition gates share a qubit, which can reduce the Schmidt decomposition of resulting three-qubit unitary to only four terms. Equivalently, every wedge reduces a cross-partition gate, and provides a speedup of a factor of 4.

2) \textit{DCD formation}. The DCD formation (see Fig.~\ref{wedge} (b)) often happens at the boundary. Specifically, the DCD formation appears when there are three successive two-qubit gates acting on the qubit pairs $(a,b)$, $(b,c)$ and $(a,b)$, and these three gates can be fused in one (the two gates on qubit pairs $(a,b)$ can be fused). A DCD formation provides a speedup of a factor of 4.

\section{II. RQC design under realistic imperfections}
\subsection{Defective qubits}

\begin{figure}[bp!]
  \includegraphics[width=0.65\columnwidth]{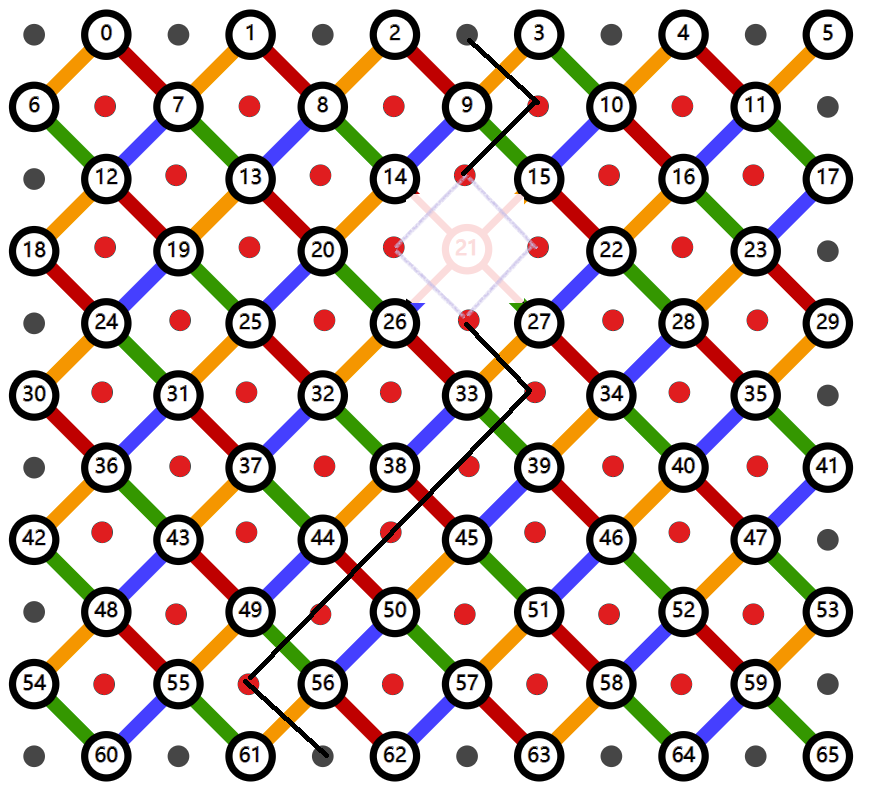} 
  \caption{\textbf{An example of RQC with one defective qubit.} The empty circles represent the qubit and the shaded empty circle represents the defective qubit. The solid circles represent the dual lattices points. The black line connecting dual lattice nodes represent a open path from one boundary site to another boundary site of the dual lattice, which is naturally a bipartition of the original lattice. The edges on the bipartition which comes across the red edges (which represent the two-qubit gate operations) are removed due to the defective qubit.
    }
    \label{fig:figSM1}
\end{figure}

Current quantum computers are prone to having defective qubits, as seen in both the Sycamore~\cite{AruteMartinisQuantumSupremacy2019} and \textit{Zuchongzhi}~\cite{WuPan2021,ZhuPan2021} quantum processors. This situation can be easily take into account with very minor changes of our method. The presence of defective qubits does not affect the procedure of exhausting all the possible RQC patterns, or the total number of different patterns. However, it could significantly affect the SFA complexity, since it will reduce the number of edges for certain bipartition which come across the defective qubits, as demonstrated in Fig.~\ref{fig:figSM1}.

\begin{figure*}[htbp!]
  \includegraphics[width=2\columnwidth]{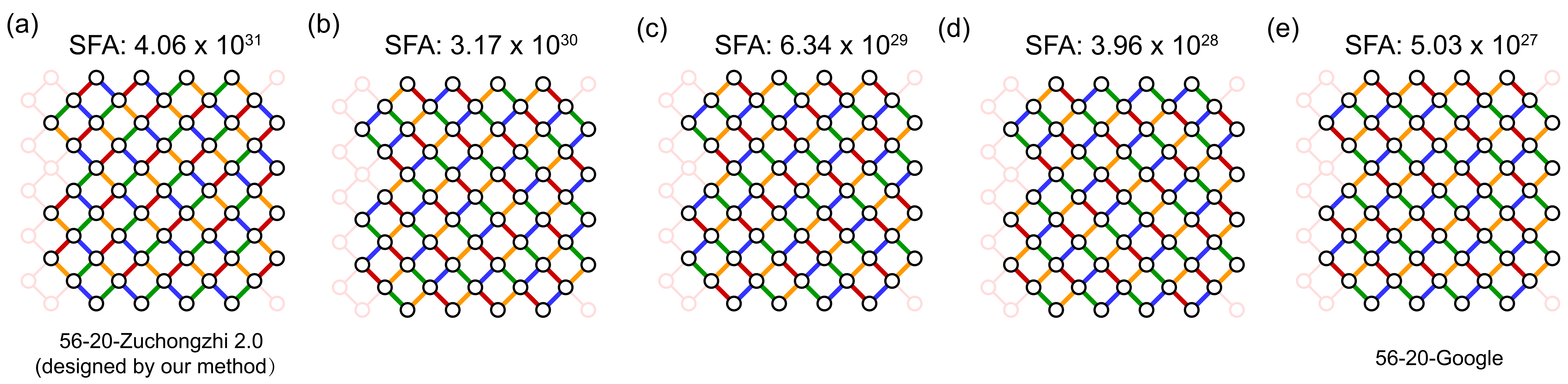} 
  \caption{\textbf{56-qubit RQCs with different SFA complexities.} 
  Each panel shows an RQC with a fixed two-qubit gate pattern, with SFA complexities decreasing from panel (a) to (e).
  The SFA complexity for RQC is shown in the head of each panel. (a) RQC designed by our method and used in the \textit{Zuchongzhi 2.0} quantum processor~\cite{WuPan2021}. (e) RQC directly adopted from the Google's experiment~\cite{AruteMartinisQuantumSupremacy2019}
    }
    \label{fig:figSM_RQC56}
\end{figure*}

To evaluate the SFA complexity of a quantum processor with defective qubits, for simplicity we reuse the same approach as done in Algorithm.1 of the main text. Namely we first pretend that the quantum processor is perfect, and search for all the valid bipartition using Algorithm.1, then we iterative over all the partitions and remove the redundant edges in those paths that have come across the defective qubits. We note that this approach have not used the defects to speedup the searching algorithm, since the speedup is likely negligible if the number of defects is very small (otherwise the classical simulation cost will be significantly reduced).

\subsection{RQC with cycles not dividable by $4$}
In certain cases one may want to extend the number of cycles such that the total number of cycles is not dividable by $4$, for example the Sycamore processor with $53$ qubits and $18$ cycles and the \textit{Zuchongzhi} processor with $60$ qubits and $22$ cycles. In these situations, simply truncating the predefined pattern $ABCDCDAB$ may not be the best choice. To design an optimal RQC in this case, we fix all the front $4\times\mod(d, 4)$ as usual, while allowing the last $r = \mod(d, 4)$ cycles to exhaust all the combinations of the four patterns. For example, if $d=18$, we take the first $16$ cycles to be the same as the hardest RQC for $d=16$, but we allow the last two cycles to exhaust all the different combinations of $A$, $B$, $C$, $D$, as long as they do not repeat (patterns such as $AA$ are not allowed since this case will certainly be much easier to classically simulate than $AB$).

\section{III. 56-qubit, 60-qubit, and 70-qubit RQCs with different SFA complexities}

\begin{figure}[htbp!]
  \includegraphics[width=\columnwidth]{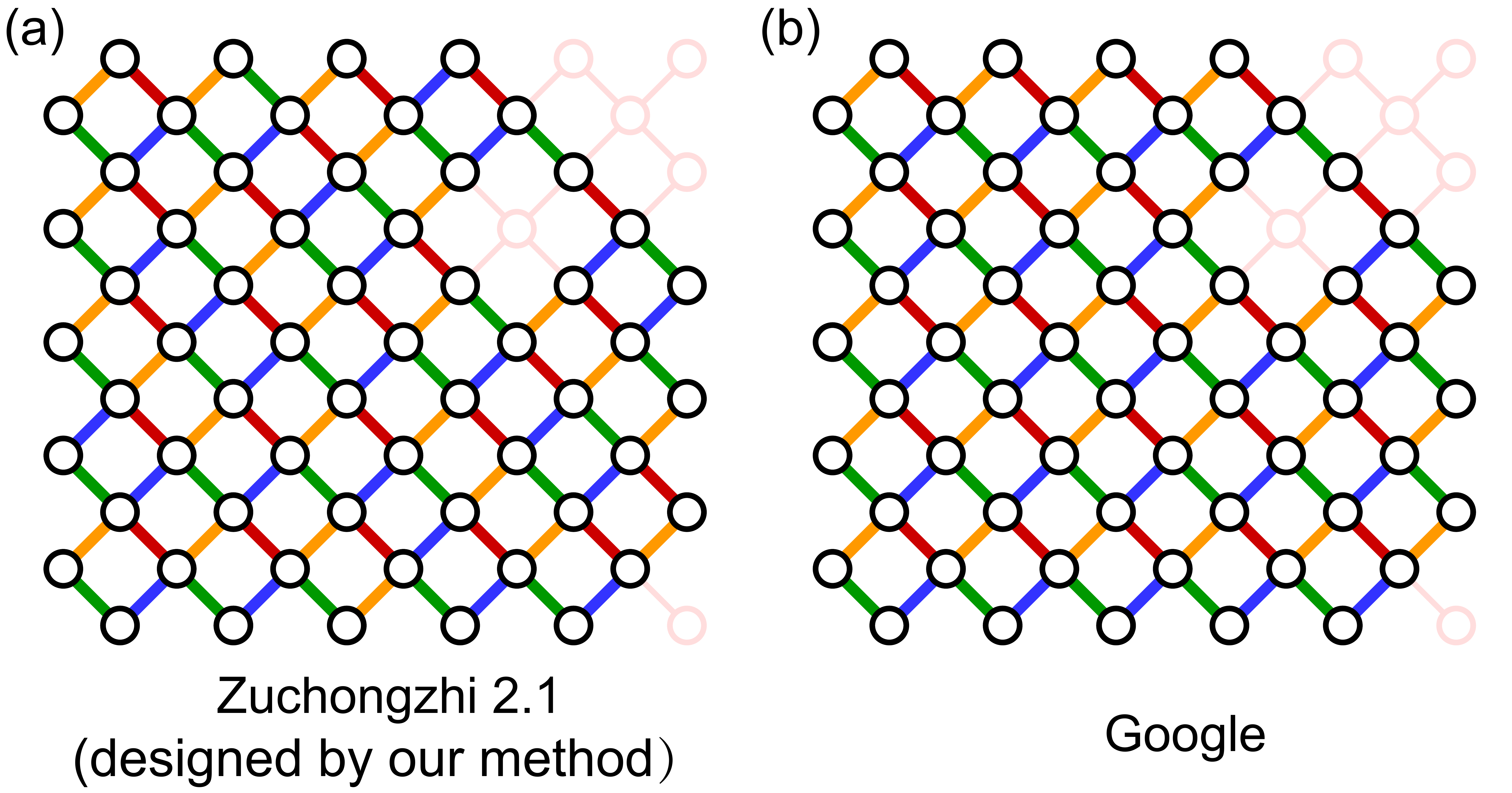} 
  \caption{\textbf{60-qubit RQC.} (a) RQC designed by our method and used in the \textit{Zuchongzhi 2.1} quantum processor~\cite{ZhuPan2021}. (b) RQC directly adopted from the Google's experiment~\cite{AruteMartinisQuantumSupremacy2019}.
    }
    \label{fig:figSM_RQC60}
\end{figure}

\begin{figure}[htbp!]
  \includegraphics[width=\columnwidth]{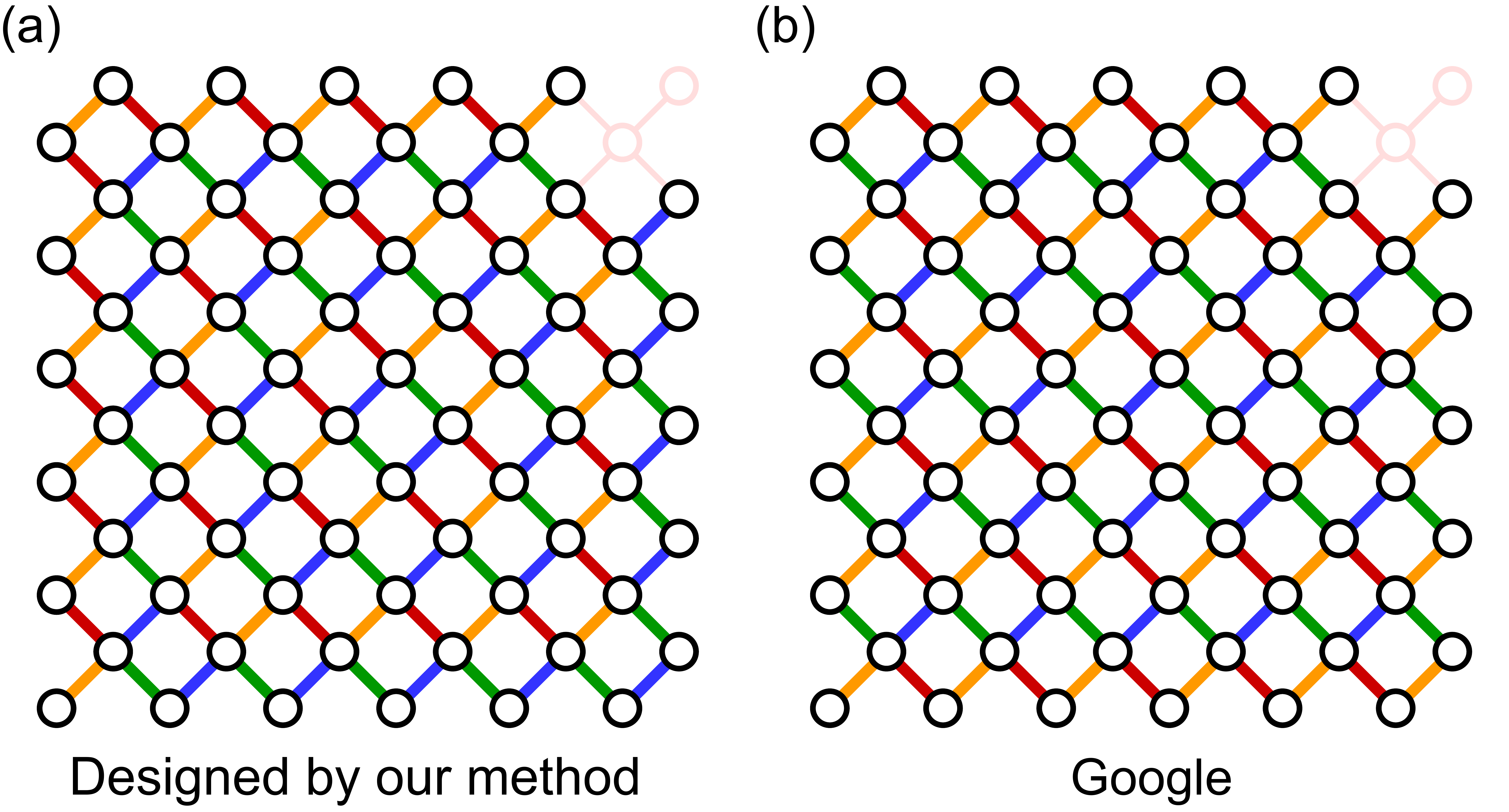} 
  \caption{\textbf{70-qubit RQC.} (a) RQC designed by our method. (b) RQC directly adopted from the Google's experiment~\cite{AruteMartinisQuantumSupremacy2019} and used in Google newest 70-qubit 24-cycle QCA experiment~\cite{Google2023b}.
    }
    \label{fig:figSM_RQC70}
\end{figure}

\begin{figure*}[!htbp]
\begin{center}
\includegraphics[width=2\columnwidth]{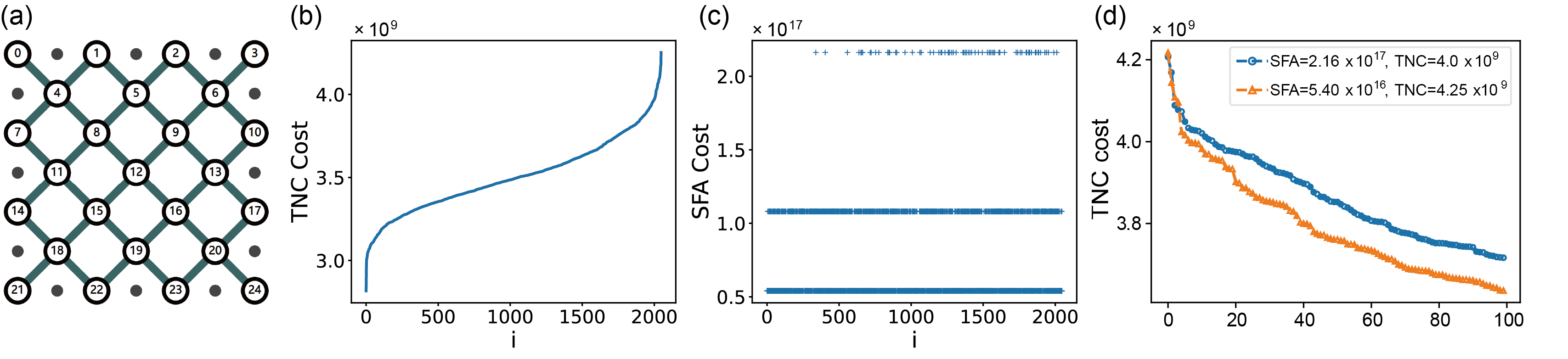}
\end{center}
\setlength{\abovecaptionskip}{0pt}
\caption{\hhl{(a) The topological structure of a 25-qubit processor, where the circles represent qubits and the edges represent possible two-qubit gate operations. (b) TNC costs for all the $2048$ possible RQCs sorted in ascending order. (c) The SFA costs for all the possible RQCs sorted in the same order as (b). (d) The estimated TNC costs with $100$ runs for the RQC with the highest TNC cost from (b) (the orange line) and for the RQC with the highest SFA cost from (c) (the blue line).}}
\label{2520TNCSFA}
\end{figure*}

In the main text, we generated RQCs with different number of qubits and circuit depths for complexity analysis. Here, we present the two-qubit gate patterns of these circuits.

Figure~\ref{fig:figSM_RQC56} provides the 56-qubit RQCs generated for Fig.2(a) in the main text. From Fig.~\ref{fig:figSM_RQC56}(a) to Fig~\ref{fig:figSM_RQC56}(e), the SFA complexity of the five 56-qubit 20-cycle RQCs gradually decreases. Among them, the RQC with the highest SFA complexity is the one used in \textit{Zuchongzhi 2.0}~\cite{WuPan2021}, which is designed by our method, while the RQC with the lowest SFA complexity is directly adopted from the Google's experiment~\cite{AruteMartinisQuantumSupremacy2019}. 

Figure~\ref{fig:figSM_RQC60} and ~\ref{fig:figSM_RQC70} provide the 60-qubit and 70-qubit RQCs generated for Fig.2(b) and  Fig.2(d) in the main text, where the RQCs in Fig.~\ref{fig:figSM_RQC60}(a) and ~\ref{fig:figSM_RQC70} (a) are designed by our method (Fig.~\ref{fig:figSM_RQC60} (a) is the one used in \textit{Zuchongzhi 2.1} experiment~\cite{ZhuPan2021}), while the RQCs in Fig.~\ref{fig:figSM_RQC60}(b) and ~\ref{fig:figSM_RQC70} (b) are directly adopted from the Google's experiment~\cite{AruteMartinisQuantumSupremacy2019}. We note that the RQC in Fig.~\ref{fig:figSM_RQC70} (b) is used in Google newest 70-qubit 24-cycle experiment~\cite{Google2023b}.

\hhl{\section{IV. RQC design using TNC}

Currently, the TNC algorithm gives the lowest classical simulation costs for RQCs. In principle, one could use the computational cost of the TNC algorithm to guide the search of the optimal RQC. However, estimating the computational cost of the TNC algorithm relys on determining a good tensor network contraction order (TNCO), and searching for the optimal TNCO is generally a hard problem. Therefore in practice the TNCO can only be determined heuristically for large tensor networks, and the resulting TNC costs may not faithfully reflect the relative complexity among different RQCs.

To illustrate this, 
we attempt to use TNC for designing an optimal 25-qubit 20-cycle RQC. There are a total of 2048 possible circuits in this case. We call the package cotengra~\cite{GrayKourtis2020} once to obtain the TNC cost for each circuit and sort them in acending order, as shown in Fig.~\ref{2520TNCSFA}(b). We also provide the corresponding SFA costs sorted in the same order as Fig.~\ref{2520TNCSFA}(c). It can be observed that the TNC costs do not correlate with the SFA costs. 

The fundamental reason is that the search for the optimal TNCO for TNC cost estimation is heuristic, 
and running this algorithm only once (or a few times) will result in huge uncertainity in determining the classical simulation complexity.
To see this, we select the RQC with the highest TNC cost from Fig.~\ref{2520TNCSFA}(b) and the RQC with the highest SFA cost from Fig.~\ref{2520TNCSFA}(c) and re-evaluated their TNC costs using cotengra for 100 times, and the results are shown in Fig.~\ref{2520TNCSFA}(d). It can be seen that the TNC cost for the RQC designed by SFA is higher than that designed by TNC. 
Therefore, to reliably determine the classical simulation cost using the TNC algorithm, one needs to run the TNCO search for a large number of times to get rid of the uncertainity, which, however, will further increase the computational costs.
As such we use the SFA cost as the guide in our approach, rather than the TNC cost.}

\hhl{\section{V. The entropy growth of RQCs with different SFA costs}

\begin{figure}[!htbp]
\begin{center}
\includegraphics[width=\columnwidth]{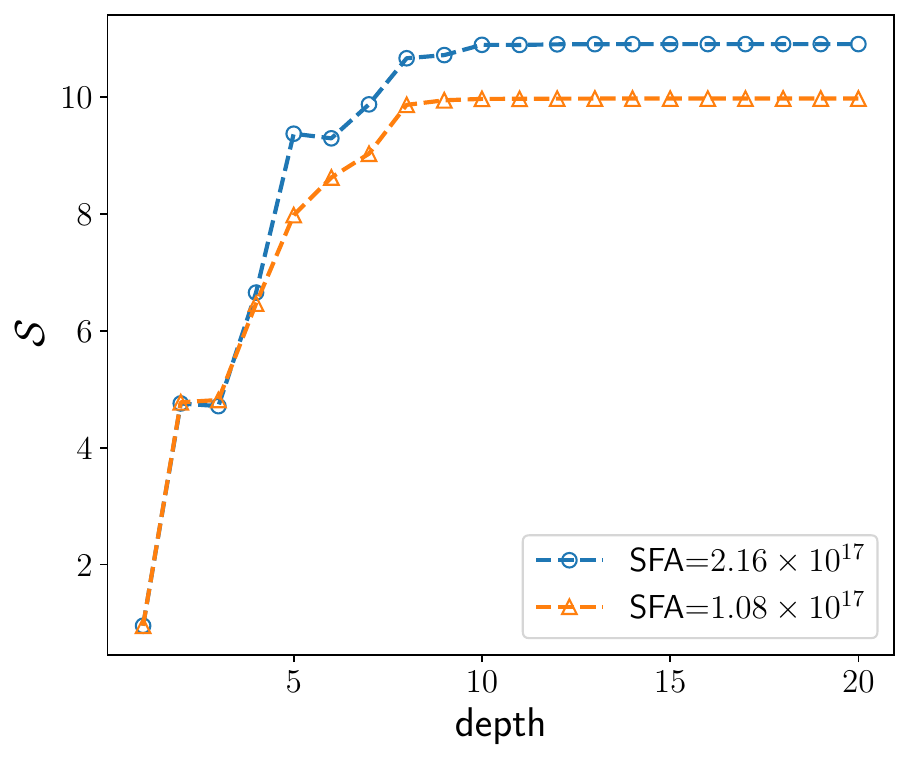}
\end{center}
\setlength{\abovecaptionskip}{0pt}
\caption{\hhl{Bipartition entanglement entropy growth as the circuit depth increases for two RQCs with different SFA costs.}}
\label{entropy}
\end{figure}

In Fig.~2(a,b,d) of the main text, we can see that generally, RQCs with high SFA costs also exhibit high TNC costs.
To gain a deeper understanding of this correlation, we calculated the entanglement entropy
$\mathcal{S} = -\sum_{\lambda} \lambda \log_2(\lambda) $ ($\lambda$s are the singular values of the reduced density matrix)
at the optimal bipartition for two 25-qubit 20-depth RQCs with different SFA costs as a function of the circuit depth (see Fig.~\ref{entropy}). It can be observed that RQCs with higher SFA costs exhibit a faster entanglement growth rate and reach a larger final value of entropy. This could be a deeper reason for the correlation of SFA and TNC costs, since the rate and magnitude of entanglement entropy growth are key factors influencing the classical simulation cost. 

}

\section{VI. The Webpage Interface}
We provide direct webpage interface~\cite{zhaohuang2021} for the second ingredient of our method, namely computing the SFA cost for a given RQC with fixed two-qubit gate patterns. While for the first ingredient of our method, namely exhausting all the possible patterns, one needs to run the source code~\cite{RQC_Github} directly. In practice, we will first determine the optimal RQCs (there will often be several RQCs with the same SFA complexity), and then use the webpage interface to visualize those RQCs to manually choose the best one (often the most symmetric one). 

\begin{figure}[!htbp]
\begin{center}
\includegraphics[width=\columnwidth]{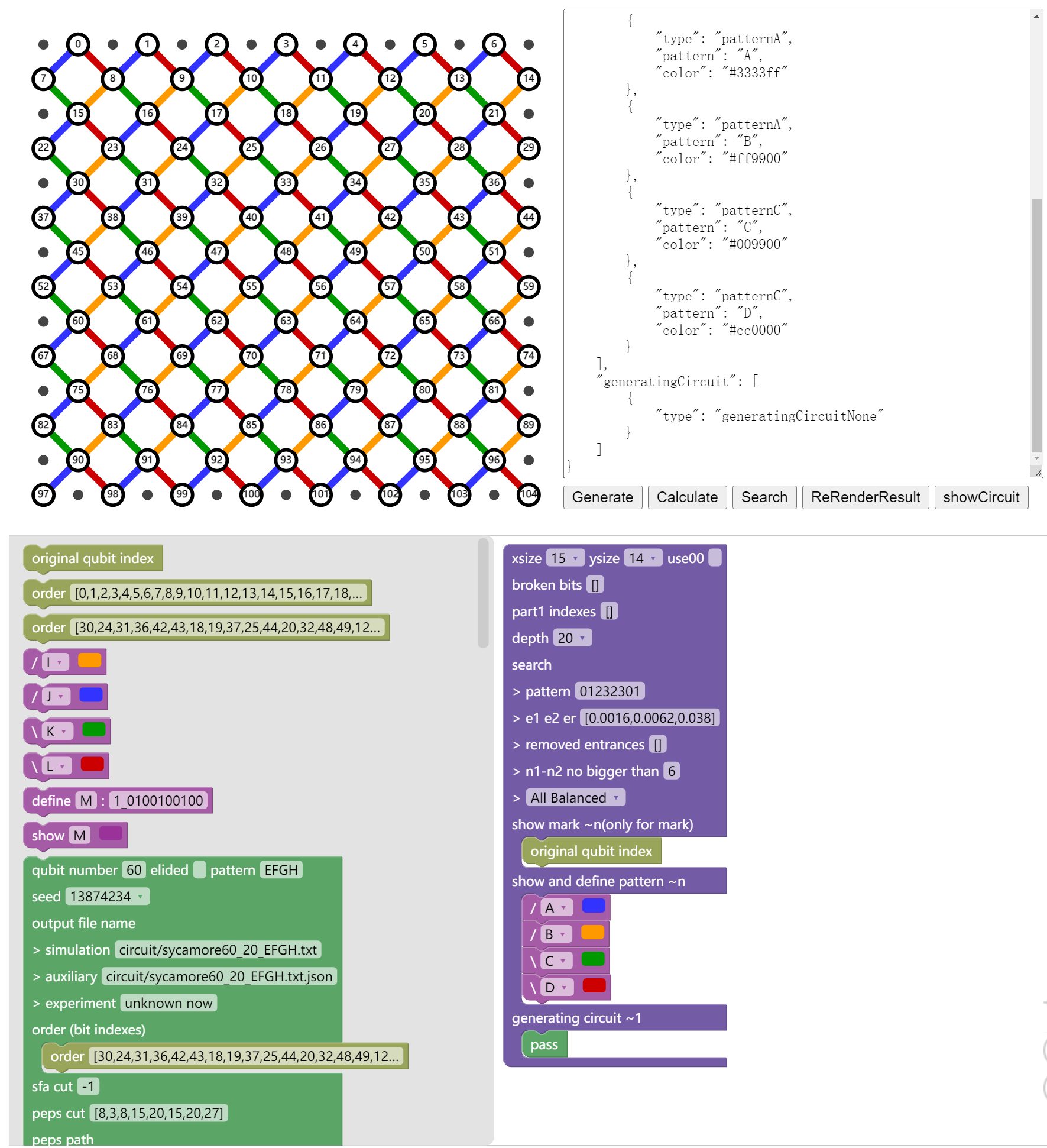}
\end{center}
\setlength{\abovecaptionskip}{0pt}
\caption{\textbf{The screenshot of the webpage interface, \textit{RQC-Cal}.}}
\label{webpage}
\end{figure}

The overview of the webpage interface contains four panels, as shown in Fig.~\ref{webpage}. The upper left panel shows the two-qubit gate patterns of the RQC, while the lower right panel contains all parameters to be specified. The upper right panel contains the source code of the input script, as well the the button to execute the program. The lower left panel contains some intermediate variables (used mainly for debugging purpose).


\begin{figure}[htbp]
\begin{center}
\includegraphics[width=\columnwidth]{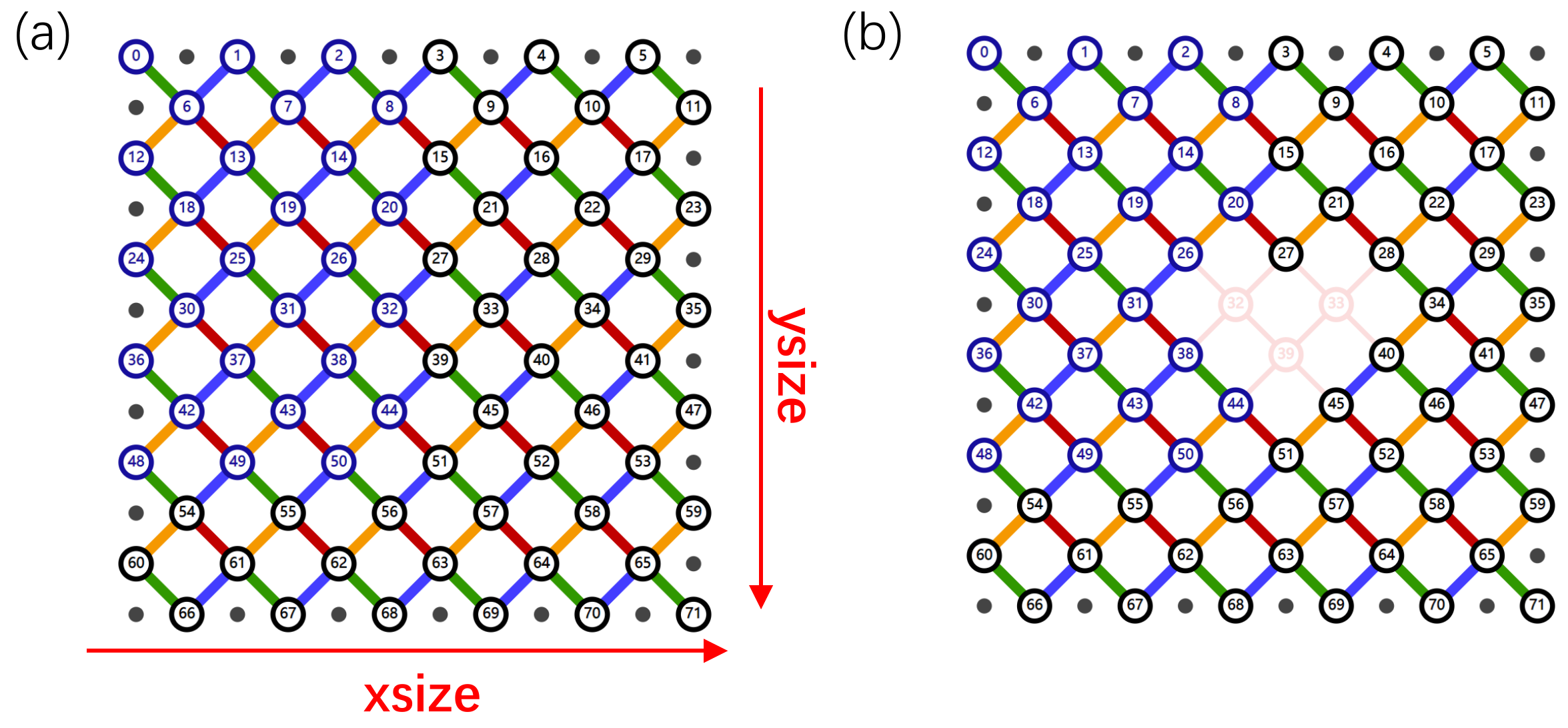}
\end{center}
\setlength{\abovecaptionskip}{0pt}
\caption{\textbf{Setting the size of the processor and defective qubits.} (a) \textit{RQC-Cal} can generate Sycamore-like processors of arbitrary sizes by setting the ``xsize'' and ``ysize'' fields.
(b) \textit{RQC-Cal} allows for flexible definition of defective qubits, such as the three defective qubits shown in the figure.}
\label{webpage-chipsize}
\end{figure}

In the next we explain the major input parameters as shown in the lower right panel of Fig.~\ref{webpage}. In the \textit{RQC-Cal}, the size of the quantum circuit (the number of qubits), and the performance parameters of noisy quantum processor can be flexibly customized. Specifically, the size of Sycamore-like quantum processor can be customized by filling in the numbers in the "xsize" and "ysize" fields of the webpage. For example, Fig.~\ref{webpage-chipsize} (a) defines a 72-qubit quantum processor with an xsize of 12 and a ysize of 12. We can also define the defective qubits in the quantum processor by typing in the broken bits field (see an example in Fig.~\ref{webpage-chipsize} (b)). The performance parameters of the quantum processor, including the fidelities of single-qubit gate, two-qubit gate, and readout, can be defined in the field "e1 e2 er".
After defining the size and performance parameters of the quantum processor, we can further input the parameters for the random quantum circuit, including the two-qubit gate patterns, the eight-cycle sequence, and the circuit depth, as indicated by the fields ``show and define pattern'', ``pattern'' and ``depth'' respectively.

\begin{figure}[htbp]
\begin{center}
\includegraphics[width=\columnwidth]{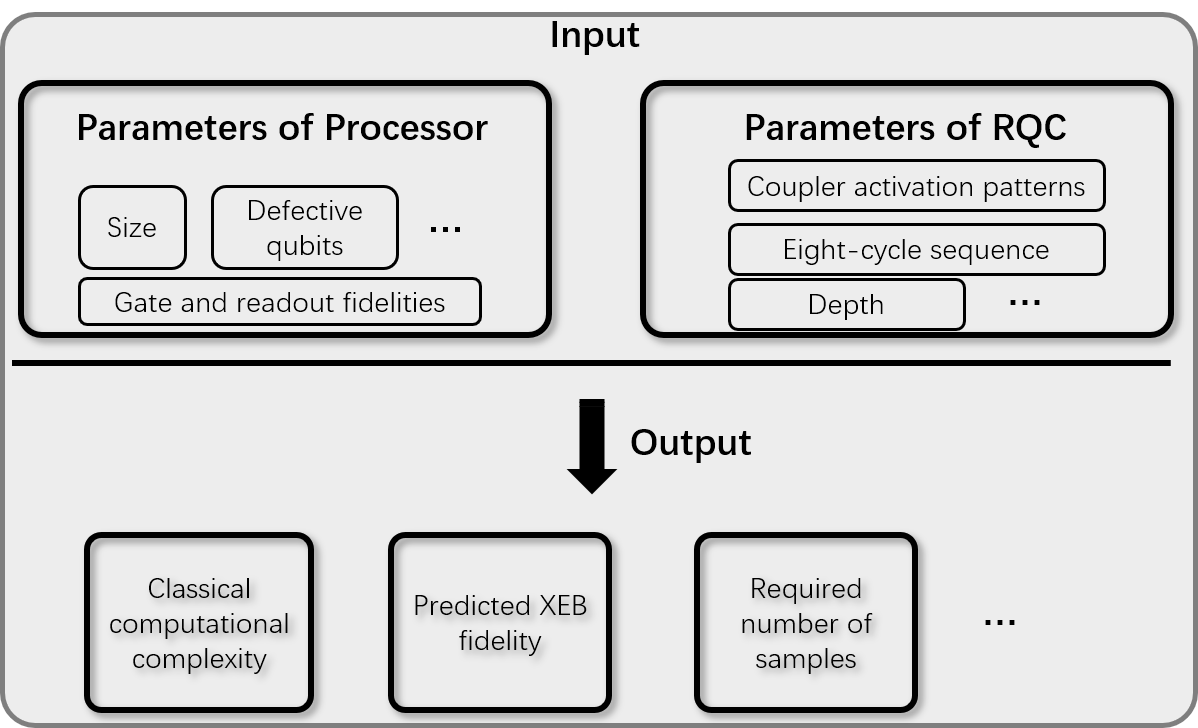}
\end{center}
\setlength{\abovecaptionskip}{0pt}
\caption{\textbf{Input and output of \textit{RQC-Cal}.} After specifying the parameters of the quantum processor (\textit{e.g.}, size, defective qubits, gate fidelity, readout fidelity, \textit{etc.}) and the parameters of the RQC (\textit{e.g.}, depth, coupler activation patterns, \textit{etc.}), \textit{RQC-Cal} will output information such as SFA computational complexity, predicted XEB fidelity, required number of samples, \textit{etc.}}
\label{webpage-IO}
\end{figure}

Given the parameters of the quantum processor and random quantum circuit, we can obtain the corresponding SFA complexity, as well as the XEB fidelity $F$ predicted by 

\begin{equation}
F{\rm{ = }}\prod\limits_{g \in {G_1}} {(1 - {e_g})} \prod\limits_{g \in {G_2}} {(1 - {e_g})} \prod\limits_{q \in Q} {(1 - {e_q})}
\label{fidelitycal}
\end{equation}
where $e_g$ are the individual gate Pauli errors, $G_1$ and $G_2$ are the set of single-qubit gates and two-qubit gates, respectively, $Q$ is the set of qubits, and $e_q$ are the state preparation and measurement errors of individual qubits. In addition, the estimated number of samples required to ensure that the 3-fold statistical error is less than the XEB fidelity, namely $N_s=\sqrt{\frac{9}{F}}$, is provided. Information about predicted XEB fidelity and required number of samples is important for the experiment because we can determine whether the experiment will be successful and how long it is expected to obtain the samples before we actually start the experiment. The information about the SFA complexity can indicate the classical simulation cost of our experiment and whether it can surpass the complexity of existing QCA experiments. The overall input-output diagram is summarized in Fig.~\ref{webpage-IO}.

\bibliography{references}